\documentclass[%
 reprint,
superscriptaddress,
nofootinbib,
 amsmath,amssymb,
 aps,
]{revtex4-2}

\usepackage{graphicx}% Include figure files
\usepackage{dcolumn}% Align table columns on decimal point
\usepackage{bm}% bold math
\usepackage{caption,subcaption}
\usepackage{tensor}
\usepackage{titlesec}
\usepackage{xcolor}
\usepackage{ytableau}
\usepackage[colorlinks,linkcolor=blue]{hyperref}
\usepackage{ytableau}
\usepackage{orcidlink}
\usepackage{academicons}
\definecolor{orcidlogocol}{HTML}{A6CE39}
\newcommand{\orcid}[1]{\href{https://orcid.org/#1}{\textcolor[HTML]{A6CE39}{\aiOrcid}}}

\usepackage{booktabs}

\begin{document}

\preprint{APS/123-QED}

\title{Gravitational-wave generation in the presence of Lorentz invariance violation}% Force line breaks with \\

\author{Samy Aoulad Lafkih \orcidlink{0009-0007-7652-8134}}
\email{samy.aoulad-lafkih@obspm.fr}
\affiliation{%
LTE, Observatoire de Paris, Université PSL, CNRS, LNE, Sorbonne Universit\'e, 61 avenue de l’Observatoire, 75014 Paris, France
}%
  \author{Marie-Christine Angonin \orcidlink{0000-0001-6000-7122}}%
 \email{marie-christine.angonin@obspm.fr}
  \author{Christophe Le Poncin-Lafitte \orcidlink{0000-0002-3811-1828}}%
 \email{christophe.leponcin-lafitte@obspm.fr}
\affiliation{%
LTE, Observatoire de Paris, Université PSL, CNRS, LNE, Sorbonne Universit\'e, 61 avenue de l’Observatoire, 75014 Paris, France
}%
\author{Nils A. Nilsson \orcidlink{0000-0001-6949-3956}}%
\email{nilsson@ibs.re.kr}%
\affiliation{Cosmology, Gravity and Astroparticle Physics Group, Center for Theoretical Physics of the Universe,
Institute for Basic Science, Daejeon 34126, Korea.}
 \affiliation{%
LTE, Observatoire de Paris, Université PSL, CNRS, LNE, Sorbonne Universit\'e, 61 avenue de l’Observatoire, 75014 Paris, France
}%

\date{\today}% It is always \today, today,

\begin{abstract}
We solve the wave equation for gravitational waves emitted by compact objects systems using the Multipolar Post-Minkowskian (MPM) method, and in the presence of Lorentz invariance violating terms. We select a Lorentz-violating extension of General Relativity in the pure gravity sector, directly taken from the Standard Model Extension (SME) formalism, and derive the wave equation for metric perturbation from the modified Einstein equation. We solve it with the MPM method and compute the gauge-invariant Riemann tensor components governing the geodesic deviation. Finally we compare the leading order term of the perturbative scheme in the small SME coefficients, with the leading order of the General Relativity. We outline the benefits and difficulties of this method. All the results are given as functionals of a set of general PM moments that can be matched to the physical properties of the source. These results are a first step toward putting state-of-the-art constraints on symmetries violations with new gravitational wave detectors like LISA.
\end{abstract}

\maketitle

%\tableofcontents

\section{Introduction}

Regions with a strong gravitational field have recently been directly probed for the first time around black holes, by the GRAVITY instrument and the Einstein Horizon Telescope collaboration. The former detected the trajectory of the star S2 at its periastron around the supermassive black hole Sagittarius A* (at the center of the Milky Way), thus constraining the relativistic precession of the orbit \cite{Grould:2017bsw, GRAVITY:2020gka, GRAVITY:2024tth}, as well as the gravitational redshift \cite{GRAVITY:2018ofz}. The latter imaged the close proximity of Sagittarius A* and of M87 (at the center of the Virgo A galaxy), offering insights into the black holes shadow, as well as the plasma structures surrounding them \cite{EventHorizonTelescope:2019dse, EventHorizonTelescope:2022wkp, EventHorizonTelescope:2024rju}.  \\
These observations are extremely useful in improving our understanding of fundamental physics, and can be complemented by other astrophysical messengers \cite{Barack:2018yly, LIGOScientific:2017ync}. For binary systems of such objects, like Neutron stars or Black Holes, they also emit gravitational waves that may be detectable in our Solar System. The study of these gravitational signals could allow us to explore the physics of these objects : the equation of state of Neutron stars \cite{Annala:2017llu}, their population astrophysical characteristics (rate of mergers can be contrasted with the existing population models) \cite{KAGRA:2021duu}, or test General Relativity \cite{Barausse:2020rsu}. \\
Gravitational waves have been theorized as early as 1916, but it took around a hundred years to detect them with the LIGO detectors (in 2015). Since then, there has been successive improvements to the instruments (current detectors are LIGO, Virgo and KAGRA) resulting in four different runs, observing around 90 events \cite{LIGOScientific:2020ibl, LIGOScientific:2021usb, KAGRA:2021vkt}. Many new instruments are planned to begin acquisition in the close future : LISA in 2035 \cite{2017arXiv170200786A}, Einstein Telescope in 2035 \cite{Maggiore:2019uih}, and Cosmic Explorers around 2037 \cite{Reitze:2019iox}. The frequency of the current Earth-based detectors is $\sim 20 - 2000$ Hz, which means that the detectable events are mainly neutron stars and black holes binary systems mergers \cite{KAGRA:2013rdx}. With its frequency range in $\sim 0.1 - 100$ mHz, LISA should be able to observe sources that differ from LIGO-Virgo-Kagra's : supermassive black hole mergers, EMRI's (Extreme Mass Ratio Inspiral), galactic binaries, and stochastic gravitational wave background \cite{Barausse:2020rsu}. The ePTA and NANOGrav collaborations have also recently reported the observation of the gravitational wave background in the correlation of the disturbance of pulsar signals \cite{EPTA:2023fyk, NANOGrav:2023gor}.\\
Up to now, General Relativity (GR) has held up exceptionally well against astrophysical tests like relativistic precessions in orbits, gravitational Doppler or light-bending \cite{Will:2014kxa}, mainly performed in zones of weak gravity. These results need to be supported by measurements in a strong gravitational field. A number of theories have already been proposed to replace General Relativity by joining it with Quantum Field Theory, and many of those invoke the breaking of the classical symmetries \cite{Horava:2009uw, Kostelecky:1988zi, Kostelecky:1991ak, Addazi:2021xuf}. The formalism known as the Standard Model Extension (SME) has specifically been introduced to offer a theory-independant way of quantifying departures from Lorentz, diffeormorphism and CPT invariance in theory, through an EFT framework \cite{Colladay:1996iz, Colladay:1998fq, Kostelecky:2000mm, Kostelecky:2002hh, Kostelecky:2003fs}. Since then, violations of these symmetries have been tested in many experiments and astrophysical observations \cite{Ding:2020aew, Hohensee:2013cya, Wolf:2006uu, Ferrari:2018tps, Stecker:2013jfa, Bourgoin:2020ckq, Pihan-LeBars:2019qsd, Hees:2015mga}, thus constraining many different subsets of the SME formalism's coefficients (see \cite{Kostelecky:2008ts} for an exhaustive list of the different constraints). Gravitational waves data have already been used to constrain the SME coefficients, but only through their dispersion and birefringence \cite{ONeal-Ault:2023lxi, ONeal-Ault:2021uwu, Haegel:2022ymk, Kostelecky:2016kfm}. The LISA instrument should allow us to place competitive constraints on the SME coefficients through the generation of gravitational waves on long-lasting events (from binary systems that are far from coalescence). Q. Bailey and al. \cite{Bailey:2023lzy, Bailey:2024zgr}, and \cite{Nilsson:2023szw} have previously obtained waveforms on subsets of the SME formalism parametrising spontaneous (\cite{Bailey:2024zgr}) and explicit (\cite{Nilsson:2023szw, Bailey:2023lzy}) spacetime symmetry breaking, using different methods : the development outlined in the reference book by E. Poisson and C. M. Will \cite{Poisson_Will_2014}, a specific coordinate transformation (previously used with the SME formalism in \cite{Bluhm:2004ep, Kostelecky:2021xhb}) and a Fourier transformation. In this paper, we will use a different method in order to expand their results : the Multipolar Post-Minkowskian (MPM) method.\\
Solving for gravitational waves is complicated due to the non-linearity of General Relativity. Many methods have been invented in order to deal with this problem, such as self-force \cite{Barack:2018yvs, Cunningham:2024dog}, amplitude scattering \cite{Damour:2017zjx, Bjerrum-Bohr:2018xdl} or the MPM method. The latter was invented by L. Blanchet and T. Damour in a series of papers in the 1980's \cite{Blanchet:1985sp, Blanchet:1986dk}. The method has accumulated numeral successes like the characterization of memory terms in the gravitational waves, up to the 4th Post-Newtonian (PN) order \cite{Trestini:2023ssa, Blanchet:1992br}. Since the method uses the PN expansion first outlined by Lorentz and Droste in 1917 and well-explained in \cite{Poisson_Will_2014}, it is fit for calculating the gravitational waves from systems that are slow-moving with respect to the speed of light, which is the case for binary systems far from coalescence. The state of the art with respect to the MPM method in General Relativity consists in the radiation-reaction force up to 4.5 PN \cite{Blanchet:2024loi} from the flux-balance equations of energy, angular and linear momentum, thus improving our understanding of how the gravitational wave's frequency evolves over time. A very complete review of the MPM method's scope and successes by L. Blanchet is available in \cite{blanchet2014gravitational}. It has also been used to study beyond-GR theories, like the scalar-tensor theory in the papers of D. Trestini \cite{Trestini:2024zpi}, or environmental effects in the case of binary system interacting with a strong electromagnetic field in \cite{Henry:2023len, Henry:2023guc}. \\
In this paper, we use the MPM method to calculate a first order correction in the small SME coefficients to the gravitational waves at the first PM order, for the model of spontaneous Lorentz invariance violations established by Q. Bailey and V. A. Kostelecký in \cite{bailey2006signals}. The method used, as well as the subset of SME coefficients explored, differentiates our approach from \cite{Nilsson:2023szw} and \cite{Bailey:2023lzy}. \\
This paper is organised as follows : in Section \ref{sec:recov}, we recall the process through which the modified Einstein equation in the weak field approximation from \cite{bailey2006signals} was obtained, and then extract a wave equation from it. Then, in Section \ref{sec:obt}, we exhibit and comment the inverse d'Alembertians from the literature before applying MPM's on our source terms, thus extracting a valid solution to our wave equation. We finish this section by adding a homogeneous solution to the previously found solution in order to make it respect the pre-imposed gauge condition. Finally, in Section \ref{sec:riem}, we compute the components of the Riemann tensor governing the geodesic deviation observed by a detector, and estimate how well such observations would be able to discriminate against the SME coefficients through a calculation of orders of magnitude.

\section{Notations}

\begin{itemize}
    \item Throughout this work, $c = 1$ (except when specified otherwise) is the speed of light and $G$ is the Newtonian gravitational constant. 
    \item We adopt the signature $(-, +, +, +)$ for the Minkowski metric $\eta_{\mu\nu}$. 
    \item We suppose that spacetime is covered by some asymptotically inertial cartesian coordinates $x^{\alpha} = (x^0, \mathbf{x})$, with $x^0 = t$ and $\mathbf{x} = (x^i)$. Greek index run from 0 to 3, and Latin index from 1 to 3. 
    \item For any tensors, the index notation $I_L$ denotes a multi-index, i.e. $I_L = I_{i_1 i_2 ... i_l}$. 
    \item We denote the Levi-Civita tensor as $\epsilon_{abc}$.
    \item Since we use the Minkowski metric to move indices up and down, starting from (\ref{eq:minkapp}) and in the rest of the article we may write spatial contracted indices on the same level for clarity's sake : $A^{k} B^{k} = A_{k} B_{k} \equiv A_{k} B^{k}$. 
    \item Notations of the form $\tensor[^{(k)}]{I}{_{\mu\nu}}$ indicates that the tensor $\tensor[]{I}{_{\mu\nu}}$ has been differentiated $k$ times with respect to $x^0$. The index of negative differentiation $^{(-1)}$ indicates a primitivation of the affixed tensor : $\tensor[^{(-1)}]{I}{_L}(u) = \displaystyle \int^{u}_{-\infty} I_L (x) \mathrm{d}x$. 
    \item Following the notations of \cite{Blanchet:1985sp} : 
    \begin{itemize}
        \item Parenthesis on the index of a tensor means a symmetrization of the tensor on those index : $I_{(L)} \equiv \frac{1}{l!} \sum_{\sigma \in T} I_{i_{\sigma(1)} i_{\sigma(2)} ... i_{\sigma(l)}}$ where $T$ is the smallest set of permutations of $(1,2,..,l)$ that makes $I_{(L)}$ fully symmetrical in $i_1, ..., i_l$. 
        \item Curly brackets on the index of a tensor means the un-normalized sum $I_{\{L\}} \equiv \sum_{\sigma \in T} I_{i_{\sigma(1)} i_{\sigma(2)} ... i_{\sigma(l)}}$.
        \item Square brackets on the index of a tensor means an anti-symmetrization on the index of the tensor through a normalized sum $I_{[L]} \equiv \frac{1}{l!} \sum_{\sigma \in T'} I_{i_{\sigma(1)} i_{\sigma(2)} ... i_{\sigma(l)}}$ where $T'$ is the smallest set of permutations of $(1,2,..,l)$ that makes $I_{[L]}$ fully anti-symmetrical in $i_1, ..., i_l$.
    \end{itemize}
\end{itemize}

\section{Recovering the wave equation}
\label{sec:recov}

In this section, we present the model from \cite{bailey2006signals, Kostelecky:2003fs} and recall the approximations through which Q. Bailey and V. A. Kostelecký have found a linearised form of the modified Einstein equations. Finally, we impose the harmonic gauge on the linearised equation in order to recover a wave equation that we analyse. 

\subsection{Modified Einstein equation}

%We consider the gravitational physics derived from the following expression of the action of gravity $S$ :\\
We consider the following action containing symmetry-breaking operators of dimension 4 :

\begin{equation}
\label{eq:Lag}
        S = \frac{1}{2 \kappa} \int \sqrt{-g} \left(R \left(1-u \right) + s^{\mu\nu} R_{\mu\nu} + t^{\lambda\kappa\mu\nu} C_{{\lambda\kappa\mu\nu}}  \right) {\rm d}^4 x + S',
\end{equation}

where $\frac{1}{2\kappa}\sqrt{-g} R$ is the classic term of General Relativity (the Lagrangian of Einstein-Hilbert), $g$ is the determinant of the metric, $\kappa = 8 \pi G$ a constant, $R$ the Ricci scalar, $R_{\mu\nu}$ the Ricci tensor, $C_{\lambda\kappa\mu\nu}$ the Weyl conformal tensor, and $S'$ is the action of any other fields (than the gravitational one) minimally coupled to the metric. 
To those curvature tensors are adjoined coefficients from the SME formalism spontaneously breaking Lorentz invariance (\cite{bailey2006signals, Kostelecky:2003fs}) : $u$, $s^{\mu\nu}$ and $t^{\lambda\kappa\mu\nu}$. As defined in \cite{bailey2006signals}, the SME tensors are traceless and possess some indices symmetry in correlation to the metric object they are contracted to : $s^{\mu\nu}$ has the index symmetries of the Ricci tensor and $t^{\lambda\kappa\mu\nu}$ of the Weyl tensor. These coefficients do transform as tensors under any transformation, at the exception of particle Lorentz transformations (see \cite{bailey2006signals, Kostelecky:2020hbb}).\\
The modified Einstein equation describing the curvature of space-time depending on the local distribution of mass and energy is obtained by varying the Lagrangian in (\ref{eq:Lag}) with respect to the metric with contravariant indices $g^{\mu\nu}$ (see \cite{bailey2006signals}),

\begin{eqnarray}
\label{eq:modEinst}
    G^{\mu\nu} - E^{\mu\nu} = \kappa (T_g)^{\mu\nu}.
\end{eqnarray}

Where $G^{\mu\nu}$ is the Einstein tensor, $(T_g)^{\mu\nu}$ is the classic energy-momentum tensor deriving from the variation of the classic matter Lagrangian, and $E^{\mu\nu}$ is the term arising from the variation of all terms involving SME coefficients in the action (\ref{eq:Lag}).\\

\begin{equation}
\label{eq:ELV}
\begin{split}
    E^{\mu\nu} =& u G^{\mu\nu} + g^{\mu\nu} \nabla^2 u - \nabla^{(\mu} \nabla^{\nu)} u \\
    &+ \frac{1}{2} s^{\alpha\beta} R_{\alpha\beta} g^{\mu\nu} - \frac{1}{2} g^{\mu \nu} \nabla_{\alpha} \nabla_{\beta} s^{\alpha\beta} \\
    &- \frac{1}{2} \nabla^{2} s^{\mu\nu} + \nabla_{\alpha} \nabla^{(\mu} s^{\nu)\alpha}  \\
    &+ t^{\alpha\beta\gamma (\mu} {R_{\alpha\beta\gamma}}^{\nu)} -2 \nabla^{\alpha} \nabla^{\beta} \tensor[]{t}{^{(\mu}_{\alpha}^{\nu)}_{\beta}} \\
    &+ \frac{1}{2} g^{\mu\nu} t^{\alpha \beta \gamma \delta} R_{\alpha \beta\gamma\delta}.
\end{split}
\end{equation}

The goal of this paper is to use this modified equation to obtain and solve the wave equation describing the gravitational waves radiated by a gravitational system (a compact objects binary, for instance).

\subsection{Perturbative expansion}

In order to obtain the wave equation defining the waveforms, the modified Einstein equation is linearised with the usual approximations (see \cite{Poisson_Will_2014}), and simplified through a series of assumptions and Taylor expansions.
First, a weak-field approximation is used to specify that the metric describes a perturbed Minkowski spacetime : 
\begin{equation}
    \label{eq:minkapp}
    \begin{split}
    g^{\mu\nu} = \eta^{\mu\nu} + h^{\mu\nu}, \quad |h^{\mu\nu}| \ll 1.
    \end{split}
\end{equation}
In the rest of the paper, any passage from contravariant to covariant index is done through contracting index with the Minkowski metric. \\
The SME coefficients are decomposed as perturbations (with a $\sim$ on top) around a background vacuum value (with a - on top) :

\begin{eqnarray}
\label{eq:Linbase1}
\begin{split}
&u = \Bar{u} + \Tilde{u}, \\
&s^{\mu\nu} = \Bar{s}^{\mu\nu} + \Tilde{s}^{\mu\nu}, \\
&t^{\kappa\lambda\mu\nu} = \Bar{t}^{\kappa\lambda\mu\nu} + \Tilde{t}^{\kappa\lambda\mu\nu}. 
\end{split}
\end{eqnarray}

In order to guarantee the translation invariance in the asymptotic Minkowski regime, as well as the conservation of the energy-momentum, the partial derivatives of the vacuum values are taken to be zero (as an assumption) in an asymptotically inertial cartesian coordinate system :

\begin{equation}
    \label{eq:Linbase2}
    \begin{split}
    &\partial_{\alpha} \Bar{u} = 0\\
    &\partial_{\alpha} \Bar{s}^{\mu\nu} = 0\\
    &\partial_{\alpha} \Bar{t}^{\kappa\lambda\mu\nu} =0\\
    \end{split}
\end{equation}

Every term of fluctuation are taken to be of the first order, and since every expressions must be of the first order for the linearisation process of (\ref{eq:modEinst}) to be respected, the terms in $\mathcal{O}(h\Tilde{u})$, $\mathcal{O}(h\Tilde{s})$, $\mathcal{O}(\Tilde{t}h)$, or $\mathcal{O}(h^2)$ have to be neglected. \\
Because the SME fields only couple with the metric (the no-coupling condition between SME fields and matter fields is an assumption of the model created in \cite{bailey2006signals}), and the leading order dynamics is governed by second derivatives of the metric, the leading order of the solution given by the equations of motion of the SME fields would be proportional to second derivatives of the metric. Therefore, it is assumed in \cite{bailey2006signals} that the SME fluctuation terms behave as second derivative of the metric contracted with the background SME field (multiplied by an unknown numerical factor).\\
Following the prescription of \cite{bailey2006signals}, some terms are introduced in order to make the linearised equation respect the Bianchi identity (null divergence) and the unknown numerical factor in the SME fluctuation terms is calibrated such that the diffeormorphism invariance is respected. Furthermore, the metric perturbation $h_{\mu\nu}$ is decomposed into the sum of $\bar{h}_{\mu\nu}$ and $\overstar{h}_{\mu\nu}$, where $\bar{h}_{\mu\nu}$ is the classic linearised gravitational-wave solutions, and $\overstar{h}_{\mu\nu}$ is a perturbation linear in the SME background coefficients $\bar{u}$ and $\bar{s}^{\mu\nu}$. Under these assumptions, one finds the expression of \cite{bailey2006signals} :

\begin{equation}
    \label{eq:LinEinst}
    \begin{split}
    \overstar{G}_{\mu\nu} =& \Bar{u} \bar{G}_{\mu\nu} + \eta_{\mu\nu} \Bar{s}^{\alpha\beta} \bar{R}_{\alpha\beta} - 2 \Bar{s}\indices{^\alpha_{(\mu}} \bar{R}_{\alpha\nu)} \\
    &+ \frac{1}{2} \Bar{s}_{\mu\nu} \Bar{R} + \Bar{s}^{\alpha\beta} \bar{R}_{\alpha\mu\nu\beta}.\\
    \end{split}
\end{equation}

Where $\overstar{G}_{\mu\nu}$ is the linearised Einstein tensor composed only of the metric perturbation $\overstar{h}_{\mu\nu}$ generated by the Lorentz violating terms. While the other curvature tensors in (\ref{eq:LinEinst}) (with a - on top) are the linearised (respectively) Einstein tensor, Ricci tensor, Ricci scalar and Riemann tensor only composed of the metric perturbation of General Relativity $\bar{h}_{\mu\nu}$.

\subsection{The wave equation}

As is usually done for this problem (see \cite{Poisson_Will_2014}), we use the trace-reverse form of the metric perturbations and we impose a gauge condition on (\ref{eq:LinEinst}) to obtain the wave equation. Taking the trace reverse of the metric perturbation allows us to simplify the field equation : $$\underline{h}_{\mu\nu} = h_{\mu\nu} - \frac{1}{2} \eta_{\mu\nu} h,$$ where $h = \eta_{\alpha\beta}h^{\alpha\beta}$ is the trace of $h^{\alpha\beta}$. This choice does not reduce the generality of the approach, as one variable can be easily obtained from the other with the identity $h = -\underline{h}$. And because General Relativity is a gauge theory with extra degrees of freedom, we may also impose the harmonic gauge condition while conserving the generality of our approach :

\begin{equation}
    \label{eq:harmonic}
    \begin{split}
    \partial^{\nu} \underline{h}_{\mu\nu} = 0\\
    \end{split}
\end{equation}

With this gauge and considering $\underline{\overstar{h}}_{\mu\nu}$ as a variable instead of $\overstar{h}_{\mu\nu}$, (\ref{eq:LinEinst}) reads :

\begin{equation}
    \label{eq:wave}
    \begin{split}
    \Box \underline{\overstar{h}}_{\mu\nu} &= \Box \left[ \Bar{u} \underline{\bar{h}}_{\mu\nu} + \eta_{\mu\nu} \Bar{s}^{\alpha\beta} \underline{\bar{h}}_{\alpha\beta} - 2 \Bar{s}\indices{^\alpha_{(\mu}} \underline{\bar{h}}_{\nu)\alpha} + \frac{1}{2} \Bar{s}_{\mu\nu} \underline{\bar{h}} \right]\\
    &- 2 \Bar{s}^{\alpha\beta} \left( \partial_{\mu} \partial_{[\nu} \bar{h}_{\beta]\alpha} + \partial_{\alpha} \partial_{[\beta} \bar{h}_{\nu]\mu} \right)
    \end{split}
\end{equation}

We use the trace-reverse of $\bar{h}_{\mu\nu}$ terms in $\Bar{s}^{\alpha\beta} \left( \partial_{\mu} \partial_{[\nu} \bar{h}_{\beta]\alpha} + \partial_{\alpha} \partial_{[\beta} \bar{h}_{\nu]\mu} \right)$ to get a form that will simplify the calculations once we have introduced the expressions of $\bar{h}_{\mu\nu}$.
We remind ourselves that the linear metric perturbations of General Relativity far from the source obey the equation : \begin{equation}
    \label{eq:wave}
    \begin{split}
    \Box \underline{\bar{h}}_{\mu\nu} = 0.
    \end{split}
\end{equation} 
We thus obtain the expression :

\begin{equation}
    \label{eq:trwave}
    \begin{split}
    \Box \underline{\overstar{h}}_{\mu\nu} =& - 2 \Bar{s}^{\alpha\beta} \left( \partial_{\mu [\nu} \underline{\bar{h}}_{\beta]\alpha} + \partial_{\alpha [\beta} \underline{\bar{h}}_{\nu]\mu} \right) \\
    &- \frac{1}{2} \left( \tensor[]{\bar{s}}{_{\nu}^{\beta}} \partial_{\mu \beta} \underline{\bar{h}} - \tensor[]{\bar{s}}{^{\alpha\beta}} \eta_{\mu\nu} \partial_{\alpha \beta} \underline{\bar{h}} + \tensor[]{\bar{s}}{^{\alpha}_{\mu}} \partial_{\alpha \nu} \underline{\bar{h}} \right) \\
    \end{split}
\end{equation}

We now introduce a solution of the linearised gravitational wave from General Relativity for $\underline{\bar{h}}_{\mu\nu}$. We consider the expansion of the Multipolar Post-Minkowskian (MPM) method (see \cite{blanchet2014gravitational}) with general undefined symmetric trace-free (STF) moments $\{I_L$, $J_L\}$ that depend only on the retarded time $u=t-r$.

\begin{equation}
    \label{eq:h0}
    \begin{split}
    \underline{\bar{h}}_{00} = -4  \sum_{l \geqslant 0} \frac{(-1)^l}{l!}& \partial^L \left( r^{-1} I_L(u) \right), \\
    \underline{\bar{h}}_{i0} = -4 \sum_{l \geqslant 1} \frac{(-1)^l}{l!}& \biggl[ \partial^{L-1} \left( r^{-1} \Dot{I}_{iL-1} (u) \right) \\
    &+ \frac{l}{l+1} \epsilon_{iab} \partial^{aL-1} \left( r^{-1} J_{bL-1} (u) \right) \biggr], \\
    \underline{\bar{h}}_{ij} = -4 \sum_{l \geqslant 2} \frac{(-1)^l}{l!}& \biggl[ \partial^{L-2} \left( r^{-1} \Ddot{I}_{ijL-2} (u) \right) \\ 
    &+ \frac{2l}{l+1} \epsilon_{ab(i} \partial^{aL-2} \left( r^{-1} \Dot{J}_{j)bL-2} (u) \right) \biggr]. \\
    \end{split}
\end{equation}

The moments $I$, $I_i$ and $J_i$ are constant (see \cite{blanchet2014gravitational}), and the dots over the moments symbolise a derivate with respect to the retarded time $u$.
The trace $\underline{\bar{h}} = -\bar{h}_{00} + \delta^{ij} \underline{\bar{h}}_{ij}$ is equal to $-\underline{\bar{h}}_{00}$ because of the antisymmetry of $\epsilon_{khl}$, as well as the trace-free property of the multipoles $\{I_L, J_L\}$.

\subsection{Defining the differential operators $\tensor[_n]{D}{_{ij}}$}

Expressing $\underline{\bar{h}}_{\mu\nu}$ as an infinite series of derivatives of the MPM moments in (\ref{eq:h0}) motivates re-writing (\ref{eq:trwave}) as a sum of differential operators $\tensor[_n]{D}{_{L}}$ acting on $\underline{\bar{h}}_{00}$, $\underline{\bar{h}}^{(0)}_{0j}$, $\underline{\bar{h}}_{ij}$ and $\underline{\bar{h}}$; in components form, the waveform reads :

\begin{equation}
    \label{eq:trwave2}
    \begin{split}
    \Box \underline{\overstar{h}}_{00} =& -\left[ \Bar{s}^{kh} \partial_{kh} \right]  \underline{\bar{h}}_{00} + 2 \left[ \Bar{s}^{kh} \partial_{0h}  \right] \underline{\bar{h}}_{0k} - \left[\Bar{s}^{kh} \partial_{00}  \right] \underline{\bar{h}}_{kh} \\ 
    &+ \frac{1}{2} \left[ \Bar{s}^{00} \partial_{00} - \Bar{s}^{kh} \partial_{kh}  \right] \underline{\bar{h}} \\
    =& \tensor[_{1}]{D}{} \underline{\bar{h}}_{00} + \tensor[_{2}]{D}{^{k}} \underline{\bar{h}}_{0k} + \tensor[_{3}]{D}{^{kh}} \underline{\bar{h}}_{kh} + \tensor[_{4}]{D}{} \underline{\bar{h}}, \\
    \\
    \Box \underline{\overstar{h}}_{0j} =& \left[ \Bar{s}^{0k} \partial_{kj} \right]  \underline{\bar{h}}_{00} \\
    &+ \left[ - \Bar{s}^{0k} \partial_{0j} - \Bar{s}^{0a} \partial_{0a} \delta_{kj} + \Bar{s}^{kl} \partial_{lj} - \Bar{s}^{al} \partial_{al} \delta_{kj} \right] \underline{\bar{h}}_{0k} \\
    &+ \left[-\Bar{s}^{kl} \partial_{0j} + \Bar{s}^{0k} \partial_{00} \delta_{lj} + \Bar{s}^{al} \partial_{0a} \delta_{kj}  \right] \underline{\bar{h}}_{kl} \\
    &+ \frac{1}{2} \left[ - \tensor[]{\Bar{s}}{_{j}^{0}} \partial_{00} - \tensor[]{\Bar{s}}{_{j}^{k}} \partial_{0k} + \Bar{s}^{00} \partial_{0j} + \Bar{s}^{k0} \partial_{kj} \right] \underline{\bar{h}} \\
    =&\tensor[_{5}]{D}{_{j}} \underline{\bar{h}}_{00} + \tensor[_{6}]{D}{^{k}_{j}} \underline{\bar{h}}_{0k} + \tensor[_{7}]{D}{^{kh}_{j}} \underline{\bar{h}}_{kh} + \tensor[_{8}]{D}{_{j}} \underline{\bar{h}}, \\
    \\
    \Box \underline{\overstar{h}}_{ij} =& -\left[ \bar{s}^{00} \partial_{ij} \right] \underline{\bar{h}}_{00} \\
    &+ 2 \left[ \bar{s}^{00} \partial_{0(i} \delta_{j)k} + \bar{s}^{0a} \partial_{a(i} \delta_{j)k} - \bar{s}^{0k} \partial_{ij} \right] \underline{\bar{h}}_{0k}  \\
    &+ \bigl[ -\left( \bar{s}^{00} \partial_{00} + 2 \bar{s}^{0a} \partial_{0a} + \bar{s}^{ab} \partial_{ab} \right) \delta_{ki} \delta_{lj} - \bar{s}^{kl} \partial_{ij} \\
    &+ 2\bar{s}^{0k} \partial_{0(i} \delta_{j)l} + 2\bar{s}^{al} \partial_{a(i} \delta_{j)k} \bigr] \underline{\bar{h}}_{kl} \\
    &+ \frac{1}{2} \Bigl[ -2 \tensor[]{\bar{s}}{^{0}_{(i}} \partial_{j)0} - 2 \tensor[]{\bar{s}}{^{k}_{(i}} \partial_{j)k} \\
    &+ \left( \bar{s}^{00} \partial_{00} + 2 \bar{s}^{0k} \partial_{0k} + \bar{s}^{kl} \partial_{kl} \right) \delta_{ij} \Bigr] \underline{\bar{h}} \\
    =&\tensor[_{9}]{D}{_{ij}} \underline{\bar{h}}_{00} + \tensor[_{10}]{D}{^{k}_{ij}} \underline{\bar{h}}_{0k} + \tensor[_{11}]{D}{^{kh}_{ij}} \underline{\bar{h}}_{kh} + \tensor[_{12}]{D}{_{ij}} \underline{\bar{h}}. 
    \end{split}
\end{equation}

Once we introduce the General Relativity solution (\ref{eq:h0}) in the source terms of (\ref{eq:trwave2}), we get source terms arranged in series. We do have to solve all three differential equations in (\ref{eq:trwave2}) because the Riemann tensor components calculated in Section \ref{sec:riem} depends on all the components of the metric : $\overstar{h}_{00}$, $\overstar{h}_{0j}$ and $\overstar{h}_{ij}$.

\section{Waveform expression}
\label{sec:obt}

We now solve the system of differential equations (\ref{eq:trwave2}) and (\ref{eq:harmonic}) for our trace-reverse (gauge-dependent) waveform. In the Section \ref{subsec:invAl} we define the inverse d'Alembertian used in Section \ref{subsec:partsol} to extract a particular solution to the wave equations (\ref{eq:trwave2}). In Section \ref{subsec:homosol}, we find and add a homogeneous solution (of the wave equation) to the particular solution found in Section \ref{subsec:partsol} in order to make the solution respect the harmonic gauge.

\subsection{Inverse d'Alembertian}
\label{subsec:invAl}

The classic way of obtaining a particular solution for a wave equation (in flat spacetime) corresponds to an integral of the retarded source term, divided by the distance between the field-point and the integration point :
%\begin{widetext}
\begin{equation}
    \label{eq:invAlemb}
    \begin{split}
    &\Box \left(F\right) (ct, \mathbf{x}) = P (ct, \mathbf{x}) \\ 
    \Longrightarrow \quad &F(ct, \mathbf{x}) = \frac{1}{4\pi} \int_{\mathbb{R}^3} \frac{P \left(ct - |\mathbf{x} - \mathbf{x'}|, \mathbf{x'} \right)}{|\mathbf{x} - \mathbf{x'}|} \mathrm{d}^3 x'.
    \end{split}
\end{equation}
%\end{widetext}

This expression for the inverse of the d'Alembertian operator allows one to view the source terms in (\ref{eq:trwave2}) in a different light. The particular solutions will probe the whole past light-cone of those GR linear gravitational waves, as if $\underline{\overstar{h}}_{\mu\nu}$ is in fact physically sourced by GR gravitational waves interacting with the SME fields in the whole space. Those interactions would produce metric perturbations that would then propagate up to the field point at the speed of light.
The formula presented in (\ref{eq:invAlemb}) is especially convenient when the source term is a linear combination of Dirac distributions, as it is the case for the energy-momentum tensor $T^{\mu\nu}$ of a system of point-like massive particles.
However, the source terms in (\ref{eq:trwave}) are different as they are double derivatives of the linear gravitational waves of General Relativity, and as such they extend in the whole space of $\mathbb{R}^3$. It was still used in \cite{Bailey:2023lzy, Nilsson:2023szw} to obtain waveforms in the presence of spacetime symmetry breaking parametrised by the SME formalism. As is prescribed in \cite{Poisson_Will_2014}, the integration space in \cite{Bailey:2023lzy, Nilsson:2023szw} is decomposed into the wave-zone and the near-zone integrals, and the former is neglected. Here we calculate both wave and near-zone integrals by considering a different expression for the inverse d'Alembertian. We use the work of L. Blanchet and T. Damour \cite{Blanchet:1985sp} to calculate the image through an inverse d'Alembertian $\Box^{-1}$ of any term of the form $\Hat{n}_L \frac{F\left( t-r \right)}{r^k}$ :
\begin{widetext}
\begin{equation}
    \label{eq:invAl}
    \begin{split}
    \Box^{-1}_R \left( \hat{n}_L r^{U-k} F(t-r) \right) &= \frac{1}{M(U-k)} \displaystyle \int_{- \infty}^{t'-r'} \mathrm{ds} F(s) \hat{\partial}'_L \left[ \frac{\left( t'-r'-s \right)^{U-k+l+2} - \left( t'+r'-s \right)^{U-k+l+2}}{r'} \right],\\
    M(U-k) &= 2^{U-k+3} (U-k+2)(U-k+1)...(U-k+2-l), \\
    \end{split}
\end{equation}
\end{widetext}
 where $l$ and $k$ are positive integers, $\hat{n}_L$ is a STF directional multipole linked to spherical harmonics (see \ref{app:form} for a reminder of the definition), and $\Hat{\partial}_L$ is an angular derivative operator whose link with the operators $\partial_L$ or $\frac{\partial .}{\partial r}$ can be found in \ref{app:form}. A source term of this form might not be properly integrable in $r=0$, but this formula takes care of such problems through a regularisation process : the source term is multiplied by $r^U$ where $U$ is a complex number whose real part satisfies the convergence needs of our integrals. This $U$ parameter is taken to be 0 at the end of the integration procedure through an analytic continuation. In this case, it is equivalent to a Hadamard regularisation procedure, which means that we merely take out the poles under the integral : in a Laurent expansion of the expression under the integral, it would be any power of $1/r$ that is higher than 2.\\

%\begin{itemize}
%    \item introduce classical inverse d'Alembertian, show difficulty
%    \item give inverse d'Alembertian in spheric with explanation of $B$
%    \item explain also STF tensor
%    \item show that our source terms are exactly suited to this inverse d'Alembertian
%    \item show calculation example with $F(s)$
%\end{itemize}

\subsection{Particular solutions}
\label{subsec:partsol}

Let us now apply the inverse d'Alembertian (\ref{eq:invAl}) to the source terms present in (\ref{eq:trwave2}). Because the commutation between $\Box^{-1}$ and any number $m$ of partial derivatives $\partial_M$ only introduces a homogeneous solution, we perform this commutation in all of $\Box^{-1} \left(\underline{\bar{h}}_{\mu\nu} \right)$ and $\Box^{-1} \left( \Box \underline{\overstar{h}}_{\mu\nu} \right)$. The particular solution reads as :

\begin{equation}
    \label{eq:trwave3}
    \begin{split}
    \Box^{-1} \left( \Box \underline{\overstar{h}}_{00} \right) =& \left( \tensor[_{1}]{D}{} - \tensor[_{4}]{D}{} \right) \Box^{-1} \left(\underline{\bar{h}}_{00} \right) \\
    &+ \tensor[_{2}]{D}{^{k}} \Box^{-1} \left(\underline{\bar{h}}_{0k} \right) + \tensor[_{3}]{D}{^{kh}} \Box^{-1} \left(\underline{\bar{h}}_{kh} \right), \\
    \\
    \Box^{-1} \left( \Box \underline{\overstar{h}}_{0j} \right) =& \left( \tensor[_{5}]{D}{_{j}} - \tensor[_{8}]{D}{_{j}} \right) \Box^{-1} \left(\underline{\bar{h}}_{00} \right) \\
    &+ \tensor[_{6}]{D}{^{k}_{j}} \Box^{-1} \left(\underline{\bar{h}}_{0k} \right) + \tensor[_{7}]{D}{^{kh}_{j}} \Box^{-1} \left(\underline{\bar{h}}_{kh} \right), \\
    \\
    \Box^{-1} \left( \Box \underline{\overstar{h}}_{ij} \right) =& \left(\tensor[_{9}]{D}{_{ij}} - \tensor[_{12}]{D}{_{ij}} \right) \Box^{-1} \left(\underline{\bar{h}}_{00}\right) \\
    &+ \tensor[_{10}]{D}{^{k}_{ij}} \Box^{-1} \left(\underline{\bar{h}}_{0k}\right) + \tensor[_{11}]{D}{^{kh}_{ij}} \Box^{-1} \left(\underline{\bar{h}}_{kh}\right), \\
    \end{split}
\end{equation}

and the image of $\underline{\bar{h}}_{\mu\nu}$ through the inverse d'Alembertian reads as :

\begin{equation}
    \label{eq:BoxInvh0}
    \begin{split}
    \Box^{-1} \left( \underline{\bar{h}}_{00} \right) \equiv& 2 \sum_{l \geqslant 0} \frac{(-1)^l}{l!} \partial^L \left(  \tensor[^{(-1)}]{I}{_L} (u) \right), \\
    \Box^{-1} \left(\underline{\bar{h}}_{i0} \right) \equiv& 2 \sum_{l \geqslant 1} \frac{(-1)^l}{l!} \biggl[ \partial^{L-1} \left( \tensor[]{I}{_{iL-1}} (u) \right) \\
    &+ \frac{l}{l+1} \epsilon_{iab} \partial^{aL-1} \left( \tensor[^{(-1)}]{J}{_{bL-1}} (u) \right) \biggr], \\
    \Box^{-1} \left(\underline{\bar{h}}_{ij} \right) \equiv& 2 \sum_{l \geqslant 2} \frac{(-1)^l}{l!} \biggl[ \partial^{L-2} \left( \tensor[]{\Dot{I}}{_{ijL-2}} (u) \right) \\
    &+ \frac{2l}{l+1} \epsilon_{ab(i} \partial^{aL-2} \left( \tensor{J}{_{j)bL-2}} (u) \right) \biggr]. \\
    \end{split}
\end{equation}

All primitivation disappear in the final result as the differential operators will always differentiate the multipoles at least once with respect to $u$.
We may also observe that, once all the derivatives are expanded, some terms that do not decrease with respect to $r$ appear. If those terms are not gauge artifacts and persist in the observables, they might be heavily constraining the SME coefficients with which they are contracted since a gravitational wave whose amplitude does not decrease with the distance between the observer and the oscillating system would appear as a very loud signal.

%\begin{itemize}
%    \item discuss some terms that disappeared in process
%    \item show the solutions
%\end{itemize}

\subsection{The homogeneous solution}
\label{subsec:homosol}

The expression (\ref{eq:BoxInvh0}) is a solution to the wave equation (\ref{eq:trwave}), but it does not necessarily respect the harmonic gauge previously imposed on $\underline{\overstar{h}}_{\mu\nu}$ (see (\ref{eq:harmonic})). We find a solution that respects both (\ref{eq:trwave}) and (\ref{eq:harmonic}) by adding a specifically chosen homogeneous solution $v_{\mu\nu}$ to (\ref{eq:BoxInvh0}). Such a solution exists because the divergence of the particular solution obtained with $\Box^{-1}$ is a vectorial homogeneous solution of the wave equation (see \ref{app:diverg}). 
If we call $\Lambda_{\mu\nu}$ the sum of the source terms in (\ref{eq:trwave}) and $\Box^{-1} \Lambda_{\mu\nu}$ the particular solution associated with the full source term (with the commutation of $\Box^{-1}$ with any partial derivatives), $v_{\mu\nu}$ must respect the conditions :

\begin{equation}
    \label{eq:homodes}
    \begin{split}
    &\Box v_{\mu\nu} = 0 \\ 
    &\partial^{\nu} \left( v_{\mu\nu} + \Box^{-1} \Lambda_{\mu\nu} \right) =0 \\ 
    &v_{[\mu\nu]} = 0 \\
    \end{split}
\end{equation}

In order to obtain $v_{\mu\nu}$, we calculate the divergence of the solution, allowing for the commutation of the inverse d'Alembertian and the partial derivatives :

\begin{equation}
    \label{eq:trwave4}
    \begin{split}
    \partial^{\nu} \Box^{-1} \Lambda_{\mu\nu} =& - \Bar{s}^{\alpha\beta} \left( \partial_{\mu} \underline{\bar{h}}_{\beta\alpha} - \partial_{\alpha} \underline{\bar{h}}_{\beta\mu} \right) - \frac{1}{2} \left( \tensor[]{\bar{s}}{^{\alpha}_{\mu}} \partial_{\alpha} \underline{\bar{h}} \right) \\
    \end{split}
\end{equation}

With the commutation of $\Box^{-1}$ and $\partial_L$, we can show that $\Box^{-1} \underline{\bar{h}}_{\mu\nu}$ is still divergenceless (see \ref{app:diverg}), and we naturally have $\Box .\Box^{-1} = id$. Because $\underline{\bar{h}}_{\mu\nu}$ is a homogeneous solution of the wave equation, the expression obtained for $\partial^{\nu} \Box^{-1} \Lambda_{\mu\nu}$ is a vectorial homogeneous solution to the wave equation.\\
Thanks to (\ref{eq:trwave4}) we find an expression satisfying (\ref{eq:homodes}) :

\begin{equation}
    \label{eq:divadd}
    \begin{split}
   v_{\mu\nu} = \bar{s}^{\alpha\beta} \eta_{\mu\nu} \underline{\bar{h}}_{\alpha \beta} - 2 \tensor[]{\bar{s}}{^{\beta}_{(\mu}} \underline{\bar{h}}_{\nu)\beta} - \frac{1}{2} \tensor[]{\bar{s}}{_{\mu\nu}} \underline{\bar{h}}_{00}.
    \end{split}
\end{equation}

We add this tensorial homogeneous solution to our particular solution (\ref{eq:trwave3}), and find the trace-reverse metric of our gravitational waves, in the harmonic gauge :

\begin{equation}
    \label{eq:fullsol}
    \begin{split}
    \underline{\overstar{h}}_{00} =& \left( \tensor[_{1}]{D}{} - \tensor[_{4}]{D}{} \right) \Box^{-1} \left(\underline{\bar{h}}_{00} \right) \\
    &+ \tensor[_{2}]{D}{^{k}} \Box^{-1} \left(\underline{\bar{h}}_{0k} \right) + \tensor[_{3}]{D}{^{kh}} \Box^{-1} \left(\underline{\bar{h}}_{kh} \right) \\
    &- \bar{s}^{\alpha\beta} \underline{\bar{h}}_{\alpha \beta} + 2 \tensor[]{\bar{s}}{^{0\beta}} \underline{\bar{h}}_{0 \beta} - \frac{1}{2} \tensor[]{\bar{s}}{^{00}} \underline{\bar{h}}_{00}, \\
    \\
    \underline{\overstar{h}}_{0j} =& \left( \tensor[_{5}]{D}{_{j}} - \tensor[_{8}]{D}{_{j}} \right) \Box^{-1} \left(\underline{\bar{h}}_{00} \right) \\
    &+ \tensor[_{6}]{D}{^{k}_{j}} \Box^{-1} \left(\underline{\bar{h}}_{0k} \right) + \tensor[_{7}]{D}{^{kh}_{j}} \Box^{-1} \left(\underline{\bar{h}}_{kh} \right) \\
    &+ \tensor[]{\bar{s}}{^{0\beta}} \underline{\bar{h}}_{j\beta} - \tensor[]{\bar{s}}{^{\beta}_{j}} \underline{\bar{h}}_{0\beta} + \frac{1}{2} \tensor[]{\bar{s}}{^{0}_{j}} \underline{\bar{h}}_{00}, \\
    \\
    \underline{\overstar{h}}_{ij} =& \left(\tensor[_{9}]{D}{_{ij}} - \tensor[_{12}]{D}{_{ij}} \right) \Box^{-1} \left(\underline{\bar{h}}_{00}\right) \\
    &+ \tensor[_{10}]{D}{^{k}_{ij}} \Box^{-1} \left(\underline{\bar{h}}_{0k}\right) + \tensor[_{11}]{D}{^{kh}_{ij}} \Box^{-1} \left(\underline{\bar{h}}_{kh}\right) \\
    &+ \delta_{ij} \bar{s}^{\alpha\beta} \underline{\bar{h}}_{\alpha \beta} - 2 \tensor[]{\bar{s}}{^{\beta}_{(i}} \underline{\bar{h}}_{j)\beta} - \frac{1}{2} \tensor[]{\bar{s}}{_{ij}} \underline{\bar{h}}_{00}. \\
    \end{split}
\end{equation}

We take the trace-reverse of this expression to find the metric perturbation $\overstar{h}_{\mu\nu}$ :

\begin{equation}
    \label{eq:fullsol2}
    \begin{split}
    \overstar{h}_{00} =& \left( \tensor[_{1}]{D}{} - \tensor[_{4}]{D}{} \right) \Box^{-1} \left(\underline{\bar{h}}_{00} \right) \\
    &+ \tensor[_{2}]{D}{^{k}} \Box^{-1} \left(\underline{\bar{h}}_{0k} \right) + \tensor[_{3}]{D}{^{kh}} \Box^{-1} \left(\underline{\bar{h}}_{kh} \right) \\
    &- \frac{1}{2} \bar{s}^{\alpha\beta} \underline{\bar{h}}_{\alpha \beta} + 2 \tensor[]{\bar{s}}{^{0\beta}} \underline{\bar{h}}_{0 \beta} - \frac{1}{2} \tensor[]{\bar{s}}{^{00}} \underline{\bar{h}}_{00}, \\
    \\
    \overstar{h}_{0j} =& \left( \tensor[_{5}]{D}{_{j}} - \tensor[_{8}]{D}{_{j}} \right) \Box^{-1} \left(\underline{\bar{h}}_{00} \right) \\
    &+ \tensor[_{6}]{D}{^{k}_{j}} \Box^{-1} \left(\underline{\bar{h}}_{0k} \right) + \tensor[_{7}]{D}{^{kh}_{j}} \Box^{-1} \left(\underline{\bar{h}}_{kh} \right) \\
    &+ \tensor[]{\bar{s}}{^{0\beta}} \underline{\bar{h}}_{j\beta} - \tensor[]{\bar{s}}{^{\beta}_{j}} \underline{\bar{h}}_{0\beta} + \frac{1}{2} \tensor[]{\bar{s}}{^{0}_{j}} \underline{\bar{h}}_{00}, \\
    \\
    \overstar{h}_{ij} =& \left(\tensor[_{9}]{D}{_{ij}} - \tensor[_{12}]{D}{_{ij}} \right) \Box^{-1} \left(\underline{\bar{h}}_{00}\right) \\
    &+ \tensor[_{10}]{D}{^{k}_{ij}} \Box^{-1} \left(\underline{\bar{h}}_{0k}\right) + \tensor[_{11}]{D}{^{kh}_{ij}} \Box^{-1} \left(\underline{\bar{h}}_{kh}\right) \\
    &+ \frac{\delta_{ij}}{2} \bar{s}^{\alpha\beta} \underline{\bar{h}}_{\alpha \beta} - 2 \tensor[]{\bar{s}}{^{\beta}_{(i}} \underline{\bar{h}}_{j)\beta} - \frac{1}{2} \tensor[]{\bar{s}}{_{ij}} \underline{\bar{h}}_{00}. \\
    \end{split}
\end{equation}

In (\ref{eq:fullsol2}) we have the full 1-PM waveform for all PN orders, but it is still in the form of a series affixed with differential operators. We have to expand all the partial derivatives, transforming them into time derivatives on the STF multipoles $\{I_L, J_L\}$. We delay this expansion until the calculation of the Riemann tensor, where we only put forward terms in powers of 0 or 1 in $1/r$, as higher powers would be drastically screened by the distance between the source and the observers. We notice that some terms are not suppressed by a strictly positive power of $1/r$ in (\ref{eq:fullsol2}). One similar behavior can be observed in the waveforms of the bumblebbee gravitational waves in \cite{Amarilo:2023wpn}, where a certain term become unsuppressed by any strictly positive powers of $1/r$ when the vacuum value vector of the bumblebee field is aligned with the position of the observer with respect to the system. \\
These unsuppressed terms in (\ref{eq:fullsol2}) seem to contradict one prime assumptions of the model in \cite{bailey2006signals}, as they mean that $h_{\mu\nu}$ does not go to zero at spatial infinity. However this can be circumvented by supposing that any astrophysical system producing these gravitational waves only started producing them at some time $\mathcal{T}$. For any $r > \mathcal{T}$, the gravitational waves are not present and the spacetime is asymptotically inertial. Mathematically, we only need for the $3^{rd}$ derivatives and above of the STF moments $\{ I_L$, $J_L \}$ to be zero when evaluated at a time $t< \mathcal{T}$.

\section{The Riemann tensor}
\label{sec:riem}

In this section, we calculate and study the components $R_{0i0j}$ of the perturbed Riemann tensor.
For gravitational waves, the observable used in detectors is the light-distance between masses at rest. In the presence of curved spacetime, the geodesic deviation is governed by spatial components of the Riemann tensor $R_{0i0j}$. We calculate those in order to obtain a gauge-independent quantity that is closely linked to the observable, and we compare the orders of magnitude of the leading-order term in the SME perturbation against the leading order term of GR.

\subsection{Riemann tensor}

When considering small metric perturbations around a Minkowski spacetine, the components of the Riemann tensor relevant to geodesic deviation are calculated with the following formula :

\begin{equation}
    \label{eq:obs1}
    \begin{split}
   2 R_{0i0j} &= \partial_{0i} h_{0j} + \partial_{0j} h_{0i} - \partial_{ij} h_{00} - \partial_{00} h_{ij}. \\
    \end{split}
\end{equation}

Calculating these components with respect to (\ref{eq:fullsol2}), one reads :

\begin{equation}
    \label{eq:Rie2}
    \begin{split}
    2 \overstar{R}_{0i0j} &= \tensor[_{1}]{A}{_{ij}} \Box^{-1} \left[\underline{\bar{h}}_{00}\right] + \tensor[_{2}]{A}{^{k}_{ij}} \Box^{-1} \left[\underline{\bar{h}}_{0k} \right] \\
    &+ \tensor[_{3}]{A}{^{kh}_{ij}} \Box^{-1} \left[\underline{\bar{h}}_{kh}\right] + \tensor[_{4}]{A}{_{ij}} \underline{\bar{h}}_{00} \\
    &+ \tensor[_{5}]{A}{^{k}_{ij}} \underline{\bar{h}}_{0k} + \tensor[_{6}]{A}{^{kh}_{ij}} \underline{\bar{h}}_{kh} , \\ 
    \end{split}
\end{equation}

where the $\tensor[_{n}]{A}{_{ij}}, \tensor[_{n}]{A}{^{k}_{ij}}, \tensor[_{n}]{A}{^{kl}_{ij}}$ ($n$ going from 1 to 6) are differential operators composed of SME coefficients and partial derivatives. In \ref{sec:appA} we define the precise expressions of the $\tensor[_{n}]{A}{_{ij}}, \tensor[_{n}]{A}{^{k}_{ij}}, \tensor[_{n}]{A}{^{kl}_{ij}}$, as well as those of the coefficients $\tensor[_{n}]{B}{^{P}_{ij}}$. The latter is not a differential operators, but a linear combination of Kronecker symbols and SME fields. In \ref{sec:appCE} we define the coefficients 
$\tensor[_{n}]{E}{^{P}_{ij}}$ and $\tensor[_{n}]{C}{^{P}_{ij}}$. These are linear combinations of the $\tensor[_{n}]{B}{^{P}_{ij}}$ coefficients, with the addition of STF directional multipoles $\hat{n}_L$.\\
These notations allow us to express the first powers of $1/r$ in $\Tilde{R}_{0i0j}$ (using formulas of \ref{app:form}) :
\begin{widetext}
\begin{equation}
    \label{eq:Rie3}
    \begin{split}
    2 \overstar{R}_{0i0j} = 2 \sum_{l \geqslant 2}& \frac{(-1)^l}{l!} \biggl[ \tensor[^{(l+3)}]{I}{_{khL-2}} \left( \sum_{1 \leqslant n \leqslant 3} \tensor[_n]{C}{^{khL-2}_{ij}} \right) + \frac{\tensor[^{(l+2)}]{I}{_{khL-2}}}{r} \biggl( \sum_{1 \leqslant n \leqslant 3} \tensor[_n]{E}{^{khL-2}_{ij}} + \sum_{4 \leqslant n \leqslant 6} \tensor[_n]{C}{^{khL-2}_{ij}} \biggr) \biggr] \\
    +& \frac{(-1)^l l}{(l+1)!} \biggl[ \epsilon_{kab} \left( \tensor[_2]{C}{^{kaL-1}_{ij}} \tensor[^{(l+3)}]{J}{_{bL-1}} + \left( \tensor[_2]{E}{^{kaL-1}_{ij}} + \tensor[_5]{C}{^{kaL-1}_{ij}} \right) \frac{\tensor[^{(l+2)}]{J}{_{bL-1}}}{r} \right) \\
    +& 2 \epsilon_{ab(k} \left( \tensor[_3]{C}{^{khaL-2}_{ij}} \tensor[^{(l+3)}]{J}{_{h)bL-2}} + \left( \tensor[_3]{E}{^{khaL-2}_{ij}} + \tensor[_6]{C}{^{khaL-2}_{ij}} \right) \frac{\tensor[^{(l+2)}]{J}{_{h)bL-2}}}{r} \right) \biggr] + \mathcal{O}\left( \frac{1}{r^2} \right).
    \end{split}
\end{equation}
\end{widetext}
We observe that there are an infinite number of STF multipoles $\{I_L$, $J_L\}$ that are at power 0 of $1/r$. Since they remain in the Riemann tensor, we know that they are not gauge artifacts. \\
The results obtained in (\ref{eq:Rie3}) represent the linear order of different perturbations : it is the 1-PM order, with linear corrections in the SME coefficients, at powers 0 and 1 of $1/r$. It does include all the PN orders of those perturbations. In order to obtain all the $n^{th}$ PN orders terms, at all PM orders, one would need to figure out which PM order contribute to the $n^{th}$ PN order, and compute those contributions. One such process is explained and shown in \cite{Blanchet:2008je} for the PM/PN orders of General Relativity. For the model used here, these calculations go beyond the scope of this article (though we do expect these STF moments to be declined as $I_L = I_L (GR) + \delta I_L + \mathcal{O}\left(\left(s^{\alpha\beta}\right)^2\right) $, where $\delta I_L = \mathcal{O}\left(s^{\alpha\beta}\right)$ and $I_L\left(GR\right)$ is the STF moment of General Relativity, with the same for $J_L$).\\
Results of the same nature have been obtained before by Q. Bailey et al. \cite{Bailey:2023lzy, Bailey:2024zgr} and in \cite{Nilsson:2023szw}, in the case of explicit and spontaneous symmetry breaking. As we do not yet have the matching of the STF moments $\{I_L, J_L\}$ to the source, we cannot explore the differences between their solutions and our term by term. However, we already know of a major difference in the presence of the non-decreasing terms. This difference could be explained by the studied subset of the SME formalism, the subset studied here entails spontaneous symmetry breaking and stems directly from \cite{bailey2006signals}, different than some of the subsets in \cite{Nilsson:2023szw, Bailey:2023lzy, Bailey:2024zgr} that study explicit symmetry breaking. It could also be explained by the method used to calculate those terms as we have used the MPM methods in this paper, whereas a more straight-forward method inspired by Poisson and Will \cite{Poisson_Will_2014} has been used in \cite{Nilsson:2023szw, Bailey:2023lzy}. The advantage of the method used in this paper is that we do not separate the domain of integration of our source terms in (\ref{eq:trwave}) in near-zone and wave-zone, but instead compute everything together. In \cite{Bailey:2024zgr}, a transformation to the momentum space is also used : some terms that were suppressed by $1/r$ but not by the small SME coefficient are obtained, whereas here the terms are suppressed by the SME coefficients but not by the distance between the source and the observer.

\subsection{Orders of magnitude}

To get the order of magnitude of the perturbation, we compare the leading terms of our ansatz (General Relativity) and our $1$st order correction due to the presence of SME coefficients. 
We compute the Riemann tensor $\bar{R}_{ij}$ (\ref{eq:obs1}) with the linear metric of General Relativity $\bar{h}_{\mu\nu}$ (\ref{eq:h0}). One reads :

\begin{equation}
    \label{eq:RieGR}
    \begin{split}
    2 \bar{R}_{0i0j} =& 2 n_k n_{(i} \left[ \frac{\tensor[^{(4)}]{I}{_{j)k}}}{r} - \frac{2}{3} \epsilon_{j)ab} n_a \frac{\tensor[^{(4)}]{J}{_{bk}}}{r} \right] \\
    &- \frac{1}{2} \left[ \frac{\tensor[^{(4)}]{I}{_{ij}}}{r} + \frac{4}{3} \epsilon_{ab(i} \frac{\tensor[^{(4)}]{J}{_{j)b}}}{r} \right] - \frac{1}{2} n_{i} n_{j} n_{k} n_{h} \frac{\tensor[^{(4)}]{I}{_{kh}}}{r}\\
    &- \frac{1}{4} \left( n_i n_j - \delta_{ij} \right)n_k n_h \frac{\tensor[^{(4)}]{I}{_{kh}}}{r} + \mathcal{O}\left(\frac{1}{r^2} \right) + \mathcal{O}\left(\frac{1}{c} \right) \\
    \end{split}
\end{equation}

In (\ref{eq:RieGR}) are only the leading orders terms in powers of $1/r$, as well as in PN orders.
The mass-quadrupole STF moment is of the order $\sim I$, and the current-quadrupole moment is $\sim J$. We find an upper bound of :  

\begin{equation}
    \label{eq:RieGRest}
    \begin{split}
    |2 \bar{R}_{0i0j} | &\lesssim \frac{3}{2} \frac{|I|}{r} \omega^4 + \frac{|J|}{r} \omega^4. \\
    \end{split}
\end{equation}

Where $\omega$ is the pulsation of the gravitational wave and corresponds to half a PN order.
For $\Tilde{R}_{0i0j}$, we take an upper bound with respect to the absolute value $|A+B| \leq |A| + |B|$ but we lower it a bit by taking all the coordinate of \textbf{n} to be zero. By doing so, we will get an expression closer to the upper bound of $|\bar{R}_{0i0j}|/|\overstar{R}_{0i0j}|$ than the ratio of the upper bounds. We also suppose that $\bar{s}^{\mu\nu}$ takes the following form :
$ \bar{s}^{\mu\nu} = 
\begin{pmatrix}
3\bar{s} & \bar{s} & \bar{s} & \bar{s}\\
\bar{s} & \bar{s} & \bar{s} & \bar{s}\\
\bar{s} & \bar{s} & \bar{s} & \bar{s}\\
\bar{s} & \bar{s} & \bar{s} & \bar{s}\\
\end{pmatrix}
$.

With these assumptions, we take the leading terms in PN orders and $1/r$ powers in (\ref{eq:Rie3}) and obtain : 

\begin{equation}
    \label{eq:RieSMEest}
    \begin{split}
    |2 \overstar{R}_{0i0j} | &\lesssim \frac{661}{231} |I| |\bar{s}| \omega^5 + \frac{2168}{315} |J| |\bar{s}| \omega^5. \\
    \end{split}
\end{equation}

Comparing expressions (\ref{eq:RieGRest}) and (\ref{eq:RieSMEest}) term by term provides us a rough estimate of the expected size of $\bar{s}^{\mu\nu}$ such that our perturbation is smaller than GR. We make the final assumption that $|I| \sim |J|$ and find :

\begin{equation}
    \label{eq:Rieratio}
    \begin{split}
    |\bar{s}| \lesssim 0.3 \times \left( \omega r \right)^{-1} \\
    \end{split}
\end{equation}

From this estimate, the SME coefficient should be inferior to the inverse of the distance between the detector and the system times half a PN order, so that the leading order component in the signal is GR.\\
This order of magnitude calculation takes into account a single signal coming from a single system. But if one considers all of the signal coming from the universe, one can realise that the sum of all the strain amplitudes would call for an extremely high signal in the detector, whatever the size of the undefined MPM moments. This seems to discriminate against this subset of SME coefficients, and as explained in \ref{app:unsupp}, against any SME subset from which a similar wave equation would arise.
 
\section{Conclusion and Discussion}

Using the model from \cite{bailey2006signals}, we calculated a first order correction in the small SME coefficients to the linearised metric of GR (\ref{eq:fullsol2}). We then plugged this metric into the components of the Riemann tensor (\ref{eq:Rie2}) governing the geodesic deviation to find a gauge-independent quantity linked to the observables. We finally expanded all partial derivatives, keeping only terms unsuppressed by $1/r$, as well as terms in $1/r$ (\ref{eq:Rie3}). By doing so we proved that this model implied the existence of terms suppressed by the SME coefficients, but not by the distance between the source and the detector. Although we leave connecting the MPM moments $\{I_L, J_L\}$ to the physical properties of the source for future work, we have compared the leading term of GR with the leading term of the perturbed metric, and we found that for the perturbation to be smaller than its ansatz (GR is the zeroth order in our perturbative scheme, where we expand with respect to a small SME coefficient), we would need some combination of the SME coefficients to be smaller than $\sim \left(\omega r \right)^{-1}$ for any system. Moreover, by considering the number of sources in the universe enclosed in the past lightcones and frequency range of current and future detectors, with respect to the unsuppressed terms, we conclude that this set of SME coefficients, as well as any other from which a similar wave equation as (\ref{eq:appunsup1}) arises, are heavily disfavored. The unsuppressed terms in the waveforms originate from the assumption that there exists a coordinate system in which the vacuum value of the SME coefficients is constant. It is explained in \cite{bailey2006signals} that the constant vacuum value is not the most general form, but a choice on the part of the authors. Though a constant vacuum value is a solution of the equations of motions of $s^{\alpha\beta}$ as well as the Einstein equations in vacuum, it seems to lead to a problematic phenomenology when it comes the generation of gravitational waves. To the authors of this paper, this is a clue that there should not exist a coordinate system in which the vacuum value is constant, though it may be very slowly varying on certain scales. \\
The analysis of the precise differences between our results and \cite{Nilsson:2023szw, Bailey:2024zgr, Bailey:2023lzy} is left for future work, but we can already note a major difference in the presence of terms unsuppressed by $1/r$ in our solution. This dissimilarity could be explained by the distinctness between the subsets of the SME formalism studied, or in the methods used (the MPM method allows one to calculate contributions from the source terms from the near-zone as well as the wave-zone). \\
In conclusion, this article presents a consistent solution to the wave equation from \cite{bailey2006signals}, with a first estimation of the constraints on the SME parameters. It needs to be completed by a matching between the MPM moments $\{I_L, J_L\}$ and the source properties to be fully consistent. Once this is done, these waveforms could be added to the LISA data pipeline so that, in the future, the SME coefficients may be estimated in the global fit \cite{Karnesis:2023ras, Littenberg:2023xpl}.

\begin{acknowledgments}
The authors thank Guillaume Faye for extremely fruitful discussions and comments. N.A.N. was financed by CNES and IBS under the project code IBS-R018-D3, and acknowledges support from PSL/Observatoire de Paris.
\end{acknowledgments}

\appendix

\section{Appendix}

\subsection{Useful formulas and definitions}
\label{app:form}

Here we include some useful definitions and properties of STF tensors from \cite{Blanchet:1985sp}.

\begin{equation}
\label{eq:app1}
    \begin{split}
        \partial_L r^{-1} = (-1)^l (2l-1)!! \frac{\hat{n}_L}{r^{l+1}} 
    \end{split}
\end{equation}

\begin{equation}
\label{eq:app2}
    \begin{split}
        \hat{n}_L = \sum^{[\frac{l}{2}]}_{k=0} (-1)^k \frac{(2l-2k-1)!!}{(2l-1)!!} \delta_{\{i_1 i_2}...\delta_{i_{2k-1} i_{2k}\}} n_{i_{2k+1}... i_{l}}
    \end{split}
\end{equation}

\begin{equation}
\label{eq:app3}
    \begin{split}
        n_L = \sum^{[\frac{l}{2}]}_{k=0} \frac{(2l-4k+1)!!}{(2l-2k+1)!!} \delta_{\{i_1 i_2}...\delta_{i_{2k-1} i_{2k}\}} \hat{n}_{i_{2k+1}... i_{l}}
    \end{split}
\end{equation}

\begin{equation}
\label{eq:app4}
    \begin{split}
        \hat{\partial}_L = \sum^{[\frac{l}{2}]}_{k=0} (-1)^k \frac{(2l-2k-1)!!}{(2l-1)!!} \delta_{\{i_1 i_2}...\delta_{i_{2k-1} i_{2k}\}} \partial_{i_{2k+1}... i_{l}} \Delta^k
    \end{split}
\end{equation}

\begin{equation}
\label{eq:app5}
    \begin{split}
        \partial_L = \sum^{[\frac{l}{2}]}_{k=0} \frac{(2l-4k+1)!!}{(2l-2k+1)!!} \delta_{\{i_1 i_2}...\delta_{i_{2k-1} i_{2k}\}} \partial_{i_{2k+1}... i_{l}} \Delta^k
    \end{split}
\end{equation}

\begin{equation}
\label{eq:app6}
    \begin{split}
        n_i \hat{n}_{a_1 a_2 ... a_l} = \hat{n}_{i a_1..a_l} + \frac{l}{2l+1} \delta_{i<a_1} \hat{n}_{a_2...a_l>}
    \end{split}
\end{equation}

\begin{equation}
\label{eq:app7}
    \begin{split}
        r \partial_i \hat{n}_L = (l+1)n_i \hat{n}_L - (2l+1) \hat{n}_{iL}
    \end{split}
\end{equation}

\begin{equation}
\label{eq:app8}
    \begin{split}
        \hat{\partial}_L r^{\lambda} = \lambda (\lambda - 2)...(\lambda -2l+2) \hat{n}_L r^{\lambda - l}, (\forall \lambda \in \mathbb{C})
    \end{split}
\end{equation}

\begin{equation}
\label{eq:app9}
    \begin{split}
        \hat{\partial}_L \left( \frac{F(t-\epsilon r)}{r} \right) &= (-\epsilon)^l \hat{n}_L \sum^{l}_{j=0} \frac{(l+j)!}{(2 \epsilon)^j j! (l-j)!} \frac{\tensor[^{(l-j)}]{F}{}(t-\epsilon r)}{r^{j+1}},\\
        &\mathrm{for} \; (\epsilon^2 = 1)
    \end{split}
\end{equation}

\subsection{Commutator $[\Box^{-1}, \partial_L]$}
\label{app:commut}

Let us show that the commutator of $\Box^{-1}$ and $\partial_L$ is a homogeneous solution of the wave equation, using the toy wave equation :

$$\Box h_L = \partial_L \Lambda,$$

where both $h_L$ and $\Lambda$ are general functions of $(ct, \mathbf{x})$. We know from the inverse d'Alembertian that $\Box^{-1} \left( \partial_L \Lambda \right)$ is a solution to the full wave equation, and because the d'Alembertian commutes with partial derivatives we have :

\begin{equation}
\label{eq:app10}
    \begin{split}
        \Box \left[ \partial_L \Box^{-1} \left( \Lambda \right) \right] = \partial_L \Lambda, \\
    \end{split}
\end{equation}

where $\partial_L \Box^{-1} \left( \Lambda \right)$ is a solution of the differential equation too, so the difference between $\partial_L \Box^{-1} \left( \Lambda \right)$ and $\Box^{-1} \left( \partial_L \Lambda \right)$ should at most be equal to a sum of homogeneous solutions.

\subsection{Divergence of the wave equation solution}
\label{app:diverg}

Thanks to \cite{Blanchet:thesis}, we know that a vectorial homogeneous solution (of the wave equation) can be mapped to a tensorial homogeneous solution such that the divergence of the tensorial solution is equal to the vectorial one.
Let us show that if the source term $\Lambda_{\mu\nu}$ of a wave equation is divergenceless, then the divergence of its image through the inverse d'Alembertian is a vectorial homogeneous solution to the wave equation. Once again, thanks to the commutation of $\Box$ and $\partial_{\alpha}$, one reads :

\begin{equation}
\label{eq:app11}
    \begin{split}
        \Box \left[ \partial^{\nu} \Box^{-1} \left( \Lambda_{\mu\nu} \right) \right] &= \partial^{\nu} \Lambda_{\mu\nu} \\
        &= 0. \\
    \end{split}
\end{equation}

\subsection{Divergence properties of $\partial_{\beta} \Box^{-1} \left( \underline{\bar{h}}_{\mu\nu} \right)$}

Let us show that for our model of $\underline{\bar{h}}_{\mu\nu}$, with the STF MPM moments $\{I_L, J_L\}$, $\Box^{-1} \left( \underline{\bar{h}}_{\mu\nu} \right)$ is a constant (when the commutation ($\Box^{-1}$, $\partial_L$) is allowed).

\begin{equation}
    \begin{aligned}
    \Box^{-1} \left[\underline{\bar{h}}_{00}\right] \equiv& - \frac{1}{2} \tensor[^{(-1)}]{I}{} + \frac{1}{2} \partial_{k} \left( \tensor[^{(-1)}]{I}{_k} \right) \\
    &- \frac{1}{2} \sum_{l\geqslant2} \frac{(-1)^l}{l!} \partial_L \left(  \tensor[^{(-1)}]{I}{_L} (u) \right), \\
    \Box^{-1} \left[\underline{\bar{h}}_{i0}\right] \equiv& \frac{1}{2} I_i - \epsilon_{iab} \partial_{a} \left( \tensor[^{(-1)}]{J}{_b}  \right) \\
    &- \frac{1}{2} \sum_{l\geqslant2} \frac{(-1)^l}{l!} \biggl[ \partial_{L-1} \Bigl( I_{iL-1}(u) \Bigr) \\
    &+ \frac{l}{l+1} \epsilon_{iab} \partial_{a L-1} \left( \tensor[^{(-1)}]{J}{_{bL-1}} (u) \right)  \biggr], \\
    \Box^{-1} \left[\underline{\bar{h}}_{ij}\right]  \equiv& - \frac{1}{2} \sum_{l\geqslant2} \frac{(-1)^l}{l!} \biggl[ \partial_{L-2} \Bigl( \Dot{I}_{ijL-2}(u) \Bigr) \\
    &+ \frac{2l}{l+1} \partial_{a L-2} \Bigl( \epsilon_{ab(i} J_{j)bL-2} (u) \Bigr)  \biggr], \\ 
    \end{aligned}
\end{equation}

This form of $\Box^{-1} \left[\underline{\bar{h}}_{\mu\nu}\right]$ implies that :

\begin{equation}
    \begin{split}
    \partial^{\nu} \Box^{-1} \left[\underline{\bar{h}}_{0\nu}\right]  &= \frac{1}{2} I, \\
    \partial^{\nu} \Box^{-1} \left[\underline{\bar{h}}_{i\nu}\right] &= - \frac{1}{2} \Dot{I}_i + \epsilon_{iab} \partial_{a} \left( \tensor[]{J}{_b}  \right) . \\
    \end{split}
\end{equation}

And $I$, $I_i$ and $J_b$ are constants, so they will disappear through the action of any differential operator.

\subsection{Derivation of a STF multipole}
\label{app:deriv}

We investigate how one can express the partial derivatives of a STF multipole $\{I_L, J_L\}$ in such a way that we can easily find the orders 0 and 1 of $(1/r)^k$.

Let $I$ be a function of only the retarded time $t-r$, and let us calculate the first powers in $1/r$ of $\partial_L \left( I \right)$ and $\partial_L \left( r^{-1} I \right)$ as : \\

\begin{equation}
    \label{eq:derI1}
    \begin{split}
    \partial_L \left( I \right) &= \sum^{[l/2]}_{k=0} c_k \delta_{\{2K} \hat{\partial}_{L-2K\}} \Delta^k I \\
    \end{split}
\end{equation}

From this we have (provable by induction) :

\begin{equation}
    \label{eq:derI2}
    \begin{split}
    \Delta^k \left( I \right) &= \partial_{0}^{2k} I - 2k \partial_{0}^{2k-1}I, \\
    \end{split}
\end{equation}

and by inserting (\ref{eq:derI2}) into (\ref{eq:derI1}), we find :

\begin{equation}
    \label{eq:derI3}
    \begin{split}
    \partial_L \left( I \right) &= \sum^{[l/2]}_{k=0} c_k \delta_{\{2K} \hat{\partial}_{L-2K\}} \left( \partial_{0}^{2k} I - 2k \partial_{0}^{2k-1} \frac{I}{r} \right) \\
    &= \sum^{[l/2]}_{k=0} c_k \delta_{\{2K} \partial^{2k}_{0} \biggl( \left[ \frac{\hat{n}_{L-2K\}}}{(-2)^{l-2k}} \sum^{l-2k}_{n=1} e_{l,k,n} r^{n-l+2k} \partial_{0}^{n}I \right] \\
    &- 2k \partial_{0}^{-1} \left[ (-1)^{l-2k} \hat{n}_{L-2K\}} \sum^{l-2k}_{n=0} f_{l,k,n} \partial_{0}^{l-2k-n} \frac{I}{r^{n+1}} \right] \biggr).\\
    \end{split}
\end{equation}
with :
$$e_{l,k,n} =  \frac{2^{n} (2l-4k-n-1)!}{(n-1)!(l-2k-n)!}$$
$$f_{l,k,n} = \frac{(l-2k+n)!}{2^n n! (l-2k-n)!}$$
We only keep the $\mathcal{O}(1/r)$ terms, so when $n=l-2k$ and $n=l-2k-1$ in the second sum, and $n=0$ in the third sum of (\ref{eq:derI3}).

\begin{equation}
    \label{eq:derI3}
    \begin{split}
    \partial_L \left( I \right) &= \sum^{[l/2]}_{k=0} c_{k,l} \delta_{\{2K} \partial^{2k}_{0} \biggl( \frac{\hat{n}_{L-2K\}}}{(-2)^{l-2k}} \Bigl[ 2^{l-2k} \partial_{0}^{l-2k}I \\
    &+ \frac{2^{l-2k-1} (l-2k)!}{(l-2k-2)!} r^{-1} \partial_{0}^{l-2k-1}I \Bigr] \\
    &- 2k \partial_{0}^{-1} \left[ (-1)^{l-2k} \hat{n}_{L-2K\}} \partial_{0}^{l-2k} \frac{I}{r} \right] \biggr) + \mathcal{O}(1/r^2).
    \end{split}
\end{equation}

Let us define $a_{k,l}$, $b_{k,l}$, $d_{k,l}$ and $c_{k,l}$ :

\begin{equation}
\begin{aligned}
    a_{k,l} =& c_{k,l} (-1)^{l}, \\
    b_{k,l} =& c_{k,l} (l-2k)(l-2k-1) \frac{(-1)^l}{2}, \\
    d_{k,l} =& 2 k c_{k,l} (-1)^{l+1}, \\
    c_{k,l} =& \frac{(2l-4k+1)!!}{(2l-2k+1)!!},
\end{aligned}
\end{equation}

and :

\begin{equation}
    \label{eq:derI4}
    \begin{split}
    \partial_L \left( I \right) =& \sum^{[l/2]}_{k=0} \delta_{\{2K} \hat{n}_{L-2K\}} \left( a_k \partial_{0}^{l}I + \left(b_k + d_k \right) \partial_{0}^{l-1} \frac{I}{r} \right)  \\
    &+ \mathcal{O}(1/r^2).
    \end{split}
\end{equation}

Let us do the same for $I/r$ :

\begin{equation}
    \label{eq:derI5}
    \begin{split}
    \partial_L \left( I/r \right) =&  \sum^{[l/2]}_{k=0} c_{k,l} \delta_{\{2K} \hat{\partial}_{L-2K\}} \partial_{0}^{2k} \left( \frac{I}{r} \right)  \\
    =& \sum^{[l/2]}_{k=0} c_k \delta_{\{2K} \hat{n}_{L-2K\}} (-1)^{l-2l} \partial_{0}^{2K} f_{l,k,0} \partial_{0}^{l-2k} \left( \frac{I}{r} \right) \\
    &+ \mathcal{O}\left( \frac{1}{r^2} \right) \\
    =& \sum^{[l/2]}_{k=0} a_k \delta_{\{2K} \hat{n}_{L-2K\}} \partial_{0}^{l} \left( \frac{I}{r} \right) + \mathcal{O} \left( \frac{1}{r^2} \right) \\
    \end{split}
\end{equation}

\subsection{Full expressions of $\tensor[_n]{A}{^L_{ij}}$ and $\tensor[_n]{B}{^L_{ij}}$}
\label{sec:appA}

\begin{equation}
    \begin{split}
    \tensor[_{1}]{A}{_{ij}} =& \frac{1}{2} \bar{s}^{\alpha\beta} \left( \delta_{ij} \partial_{00\alpha\beta} + \partial_{ij\alpha\beta} \right) \nonumber \\
    =& \frac{1}{2} \Bigl( \bar{s}^{00} \delta_{ij} \partial_{0}^4 + 2 \bar{s}^{0a} \delta_{ij} \partial_{000a} + \left(\bar{s}^{ab} \delta_{ij} + s^{00} \delta_{ai} \delta_{bj} \right) \partial_{00ab} \nonumber \\
    &+ 2 \bar{s}^{0a} \delta_{bi} \delta_{cj} \partial_{0abc} + \bar{s}^{ab} \delta_{bi} \delta_{cj} \partial_{abcd} \Bigr) \nonumber \\
    =& \sum_{0 \leqslant p \leqslant 4} \tensor[_1]{B}{^{P}_{ij}} \partial_{0}^{4-p} \partial_{P}, \nonumber \\
    \\
    \tensor[_{2}]{A}{^{k}_{ij}} =& -2 \bar{s}^{\alpha\beta} \partial_{\alpha\beta0(i} \delta_{j)k}  \nonumber \\
    =& -2 \delta_{k(i} \delta_{j)c} \left( \bar{s}^{00} \partial_{000c} + 2 \bar{s}^{0a} \partial_{00ac} + \bar{s}^{ab} \partial_{abc0} \right) \nonumber \\
     =& \sum_{1 \leqslant p \leqslant 3} \tensor[_2]{B}{^{kP}_{ij}} \partial_{0}^{4-p} \partial_{P}, \nonumber \\
     \\
     \tensor[_{3}]{A}{^{kh}_{ij}} =& \bar{s}^{\alpha\beta} \partial_{00 \alpha\beta} \delta_{ki} \delta_{hj} \nonumber \\
    =& \delta_{ki} \delta_{hj} \left( \bar{s}^{00} \partial_{0}^4 + 2 \bar{s}^{0a} \partial_{000a} + \bar{s}^{ab} \partial_{00ab} \right) \nonumber \\
    =& \sum_{0 \leqslant p \leqslant 2} \tensor[_3]{B}{^{khP}_{ij}} \partial_{0}^{4-p} \partial_{P}, \nonumber \\
    \\
    \tensor[_{4}]{A}{_{ij}} =& - \bar{s}^{00} \left( \frac{1}{2} \delta_{ij} \partial_{00} + \partial_{ij} \right) - \tensor[]{\bar{s}}{^{0}_{(i}} \partial_{j)0} + \frac{1}{2} \bar{s}_{ij} \partial_{00} \\
    =& \frac{1}{2} \left( \bar{s}_{ij} - \bar{s}^{00} \delta_{ij} \right) \partial_{00} - \tensor[]{\bar{s}}{^{0}_{(i}} \delta_{j)a} \partial_{0a} - \bar{s}^{00} \delta_{ai} \delta_{bj} \partial_{ab} \\
    =& \sum_{0 \leqslant p \leqslant 2} \tensor[_4]{B}{^{P}_{ij}} \partial_{0}^{2-p} \partial_{P}, \nonumber \\
    \end{split}
\end{equation}

\begin{equation}
\label{eq:opA}
    \begin{split}
    \tensor[_{5}]{A}{^{k}_{ij}} =& 2 \bar{s}^{00} \partial_{0(i} \delta_{j)k} - \bar{s}^{0k} \left( \partial_{ij} + \delta_{ij} \partial_{00} \right) + 2 \tensor[]{\bar{s}}{^{0}_{(i}} \delta_{j)k} \partial_{00} \\
    &- 2 \tensor[]{\bar{s}}{^{k}_{(i}} \partial_{j)0} \\
    =& \left( 2 \tensor[]{\bar{s}}{^{0}_{(i}} \delta_{j)k} - \bar{s}^{0k} \delta_{ij} \right) \partial_{00} \\
    &+ 2 \left( \bar{s}^{00} \delta_{k(i} - \tensor[]{\bar{s}}{^{k}_{(i}} \right) \delta_{j)a} \partial_{0a} - \bar{s}^{0k} \delta_{ai} \delta_{bj} \partial_{ab} \\ 
    =& \sum_{0 \leqslant p \leqslant 2} \tensor[_5]{B}{^{kP}_{ij}} \partial_{0}^{2-p} \partial_{P}, \\
    \\
    \tensor[_{6}]{A}{^{kh}_{ij}} =& 2 \bar{s}^{0k} \partial_{0(i} \delta_{j)h} + 2 \tensor[]{\bar{s}}{^{k}_{(i}} \delta_{j)h} \partial_{00} + \frac{1}{2} \bar{s}^{kh} \left( \partial_{ij} - \delta_{ij} \partial_{00} \right) \\
    =& \left(2 \tensor[]{\bar{s}}{^{k}_{(i}} \delta_{j)h} - \frac{1}{2} \bar{s}^{kh} \delta_{ij} \right) \partial_{00} + 2 \bar{s}^{0k} \delta_{h(i} \delta_{j)a} \partial_{0a} \\
    &+ \frac{1}{2} \bar{s}^{kh} \delta_{ai} \delta_{bj} \partial_{ab} \\
    =& \sum_{0 \leqslant p \leqslant 2} \tensor[_6]{B}{^{khP}_{ij}} \partial_{0}^{2-p} \partial_{P}. \\
    \end{split}
\end{equation}

\subsection{Coefficients $\tensor[_n]{C}{^L_{ij}}$ and $\tensor[_n]{E}{^L_{ij}}$}
\label{sec:appCE}

\begin{equation}
    \begin{split}
    &\tensor[_{1}]{C}{^L_{ij}} = \sum_{0 \leqslant p \leqslant 4} \tensor[_1]{B}{^{P}_{ij}} \sum^{[\frac{p+l}{2}]}_{q=0} \delta_{\{2Q} \hat{n}_{L+P-2Q\}} a_{q,l}, \\
    &\tensor[_{1}]{E}{^L_{ij}} = \sum_{0 \leqslant p \leqslant 4} \tensor[_1]{B}{^{P}_{ij}} \sum^{[\frac{p+l}{2}]}_{q=0} \delta_{\{2Q} \hat{n}_{L+P-2Q\}} \left(b_{q,l} + d_{q,l} \right) \nonumber, \\
    \\
    &\tensor[_{2}]{C}{^{kL}_{ij}} = \sum_{1 \leqslant p \leqslant 3} \tensor[_2]{B}{^{kP}_{ij}} \sum^{[\frac{p+l}{2}]}_{q=0} \delta_{\{2Q} \hat{n}_{L+P-2Q\}} a_{q,l}, \nonumber \\
    &\tensor[_{2}]{E}{^{kL}_{ij}} = \sum_{1 \leqslant p \leqslant 3} \tensor[_2]{B}{^{kP}_{ij}} \sum^{[\frac{p+l}{2}]}_{q=0} \delta_{\{2Q} \hat{n}_{L+P-2Q\}} \left(b_{q,l} + d_{q,l} \right), \nonumber \\
    \\
    &\tensor[_{3}]{C}{^{khL}_{ij}} = \sum_{0 \leqslant p \leqslant 2} \tensor[_3]{B}{^{klP}_{ij}} \sum^{[\frac{p+l}{2}]}_{q=0} \delta_{\{2Q} \hat{n}_{L+P-2Q\}} a_{q,l}, \nonumber \\
    &\tensor[_{3}]{E}{^{khL}_{ij}} = \sum_{0 \leqslant p \leqslant 2} \tensor[_3]{B}{^{klP}_{ij}} \sum^{[\frac{p+l}{2}]}_{q=0} \delta_{\{2Q} \hat{n}_{L+P-2Q\}} \left(b_{q,l} + d_{q,l} \right),  \nonumber\\
    \end{split}
\end{equation}

\begin{equation}
\label{eq:opC}
    \begin{split}
        &\tensor[_{4}]{C}{^L_{ij}} = \sum_{0 \leqslant p \leqslant 2} \tensor[_4]{B}{^{P}_{ij}} \sum^{[\frac{p+l}{2}]}_{q=0} \delta_{\{2Q} \hat{n}_{L+P-2Q\}} a_{q,l}, \\
        &\tensor[_{4}]{E}{^L_{ij}} = \sum_{0 \leqslant p \leqslant 2} \tensor[_4]{B}{^{P}_{ij}} \sum^{[\frac{p+l}{2}]}_{q=0} \delta_{\{2Q} \hat{n}_{L+P-2Q\}} \left(b_{q,l} + d_{q,l} \right),  \\
        \\
        &\tensor[_{5}]{C}{^{kL}_{ij}} = \sum_{0 \leqslant p \leqslant 2} \tensor[_5]{B}{^{kP}_{ij}} \sum^{[\frac{p+l}{2}]}_{q=0} \delta_{\{2Q} \hat{n}_{L+P-2Q\}} a_{q,l}, \\
        &\tensor[_{5}]{E}{^{kL}_{ij}} = \sum_{0 \leqslant p \leqslant 2} \tensor[_5]{B}{^{kP}_{ij}} \sum^{[\frac{p+l}{2}]}_{q=0} \delta_{\{2Q} \hat{n}_{L+P-2Q\}} \left(b_{q,l} + d_{q,l} \right), \\
        \\
        &\tensor[_{6}]{C}{^{khL}_{ij}} = \sum_{0 \leqslant p \leqslant 2} \tensor[_6]{B}{^{khP}_{ij}} \sum^{[\frac{p+l}{2}]}_{q=0} \delta_{\{2Q} \hat{n}_{L+P-2Q\}} a_{q,l}, \\
        &\tensor[_{6}]{E}{^{khL}_{ij}} = \sum_{0 \leqslant p \leqslant 2} \tensor[_6]{B}{^{khP}_{ij}} \sum^{[\frac{p+l}{2}]}_{q=0} \delta_{\{2Q} \hat{n}_{L+P-2Q\}} \left(b_{q,l} + d_{q,l} \right), \\
    \end{split}
\end{equation}

where the sequences $a_{q,l}$, $b_{q,l}$ and $d_{q,l}$ have the following expressions :

$$a_{q,l} = c_{q,l} (-1)^{l}$$
$$b_{q,l} = c_{q,l} (l-2q)(l-2q-1) \frac{(-1)^l}{2}$$
$$d_{q,l} = 2 q c_{q,l} (-1)^{l+1}$$
$$c_{q,l} = \frac{(2l-4q+1)!!}{(2l-2q+1)!!}$$

\subsection{Unsuppressed terms}
\label{app:unsupp}

The terms unsuppressed by $1/r$ in (\ref{eq:fullsol}) come from a specific form of the source terms. Any gravitation theory whose linearised gravitational waves are described by an equation of the sort : 

\begin{equation}
    \label{eq:appunsup1}
    \begin{split}
    \Box \overstar{h}_{\mu\nu} = \sum_{l} \tensor[]{K}{^{\alpha \beta \epsilon_1 ... \epsilon_l}_{\mu\nu}} \partial_{\epsilon_1 ... \epsilon_l} \underline{h}_{\alpha_\beta}. \\
    \end{split}
\end{equation}

Where $\overstar{h}_{\mu\nu}$ are beyond-GR metric perturbations, as in (\ref{eq:wave}) $\underline{h}_{\alpha\beta}$ is the 1-PM metric perturbation from General Relativity, and the $\tensor[]{K}{^{\alpha \beta \epsilon_1 ... \epsilon_l}}$ are tensors that respect the identity $\partial_{\gamma} \tensor[]{K}{^{\alpha \beta \epsilon_1 ... \epsilon_l}_{\mu\nu}} = 0$.
Under these conditions, the solution of (\ref{eq:appunsup1}), up to a homogeneous solution, is :

\begin{equation}
    \label{eq:appunsup2}
    \begin{split}
    \overstar{h}_{\mu\nu} = \sum_{l} \tensor[]{K}{^{\alpha \beta \epsilon_1 ... \epsilon_l}_{\mu\nu}} \partial_{\epsilon_1 ... \epsilon_l} \Box^{-1} \underline{h}_{\alpha_\beta}. \\
    \end{split}
\end{equation}

With $\Box^{-1} \underline{h}_{\alpha_\beta}$ defined by (\ref{eq:BoxInvh0}). One can see immediately that a lot of terms unsuppressed by $1/r$ and proportional to the time derivated multipoles \{$I_L$, $J_L$\} appear. And since these unsuppressed terms are not homogeneous solutions of the wave equations, they cannot be negated by adding a carefully chosen homogeneous solution. \\
Additionnaly, if one considers the gauge-moment laid-out in \cite{blanchet2014gravitational}, they cannot be absorbed in a gauge transformation as well. Specifically, it means that any SME formalism subset with a vacuum value (for spontaneous breaking), or an explicit value (for explicit breaking), constant everywhere in some coordinate system would most likely suffer from the same pathology. For instance, the linearised equations in \cite{Kostelecky:2016kfm, Kostelecky:2017zob} would give rise to the same unsuppressed terms. One can apply the insights of this article to those cases and conclude that those SME subsets are heavily disfavored as well. 

%\nocite{*}

\bibliography{SME_GW/SME_GW}% Produces the bibliography via BibTeX.

\end{document}